\documentclass[]{mn2e}

\input{psfig.sty}
\def\gs{\mathrel{\raise0.35ex\hbox{$\scriptstyle >$}\kern-0.6em
\lower0.40ex\hbox{{$\scriptstyle \sim$}}}}
\def\ls{\mathrel{\raise0.35ex\hbox{$\scriptstyle <$}\kern-0.6em
\lower0.40ex\hbox{{$\scriptstyle \sim$}}}}

\makeatother

\title[X-ray emission from TN\,J1338$-$1942 at $z= $\,4.1]{
X-ray emission around the \emph{z}\,=\,4.1 radio galaxy TN\,J1338$-$1942 and the potential role of far-infrared photons in AGN Feedback
}

\author[Smail \& Blundell]{Ian Smail$^{1}$\thanks{ian.smail@durham.ac.uk} \& Katherine M.\ Blundell$^{2}$\\\\
$^{1}$Institute for Computational Cosmology, Durham University, South Road,
        Durham DH1 3LE UK\\
$^{2}$University of Oxford, Astrophysics, Keble Road, Oxford OX1 3RH UK\\
}

\begin{document}

\pagerange{1--6} \pubyear{2013}
\volume{666}

\maketitle 

\begin{abstract}
We report the discovery in an 80-ks observation of spatially-extended X-ray emission around the high-redshift radio galaxy TN\,J1388$-$1942 ($z=$\,4.11) with the {\it Chandra X-ray Observatory}.  The X-ray emission extends over a $\sim $\,30-kpc diameter region and although it is less extended than the GHz-radio lobes, it is roughly aligned with them. We suggest that the X-ray emission arises from Inverse Compton (IC) scattering of photons by relativistic electrons around the radio galaxy.  At $z=$\,4.11 this is the highest redshift detection of IC emission around a radio galaxy.  We investigate the hypothesis that in this compact source, the Cosmic Microwave Background (CMB), which is $\sim$\,700$\times$ more intense than at $z\sim$\,0 is nonetheless not the relevant seed photon field for the bulk of the IC emission.  Instead, we find a tentative correlation between the IC emission and far-infrared luminosities of compact, far-infrared luminous high-redshift radio galaxies (those with lobe lengths of $\ls $\,100\,kpc). Based on these results we suggest that in the earliest phases of the evolution of radio-loud AGN at very high redshift, the far-infrared photons from the co-eval dusty starbursts occuring within these systems may make a significant contribution to their IC X-ray emission  and so contribute to the feedback in these massive high-redshift galaxies. 
\end{abstract}


%
%
\begin{figure*}
  \centerline{\psfig{file=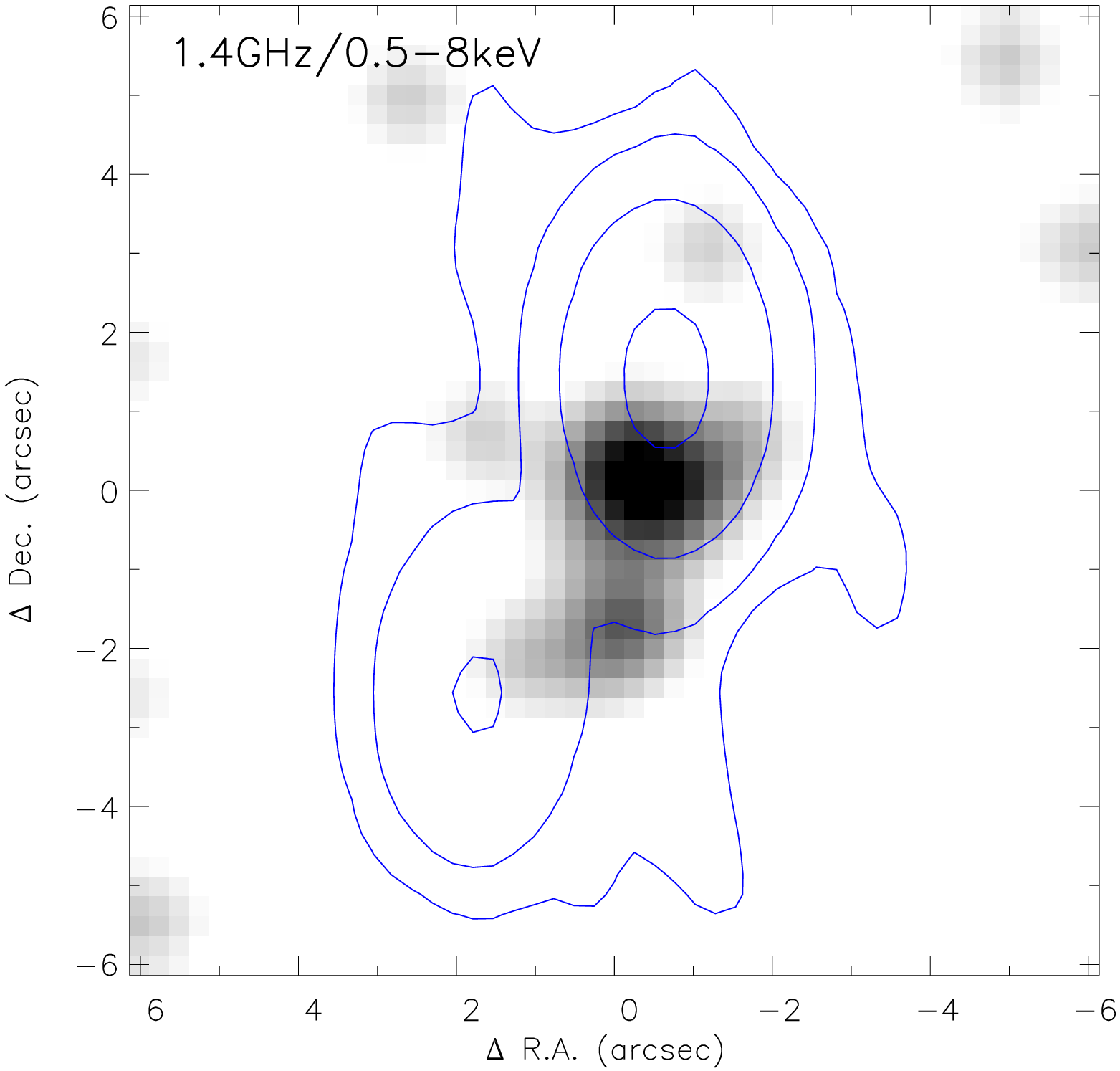,width=1.9in,angle=0}\hspace*{0.2cm}
    \psfig{file=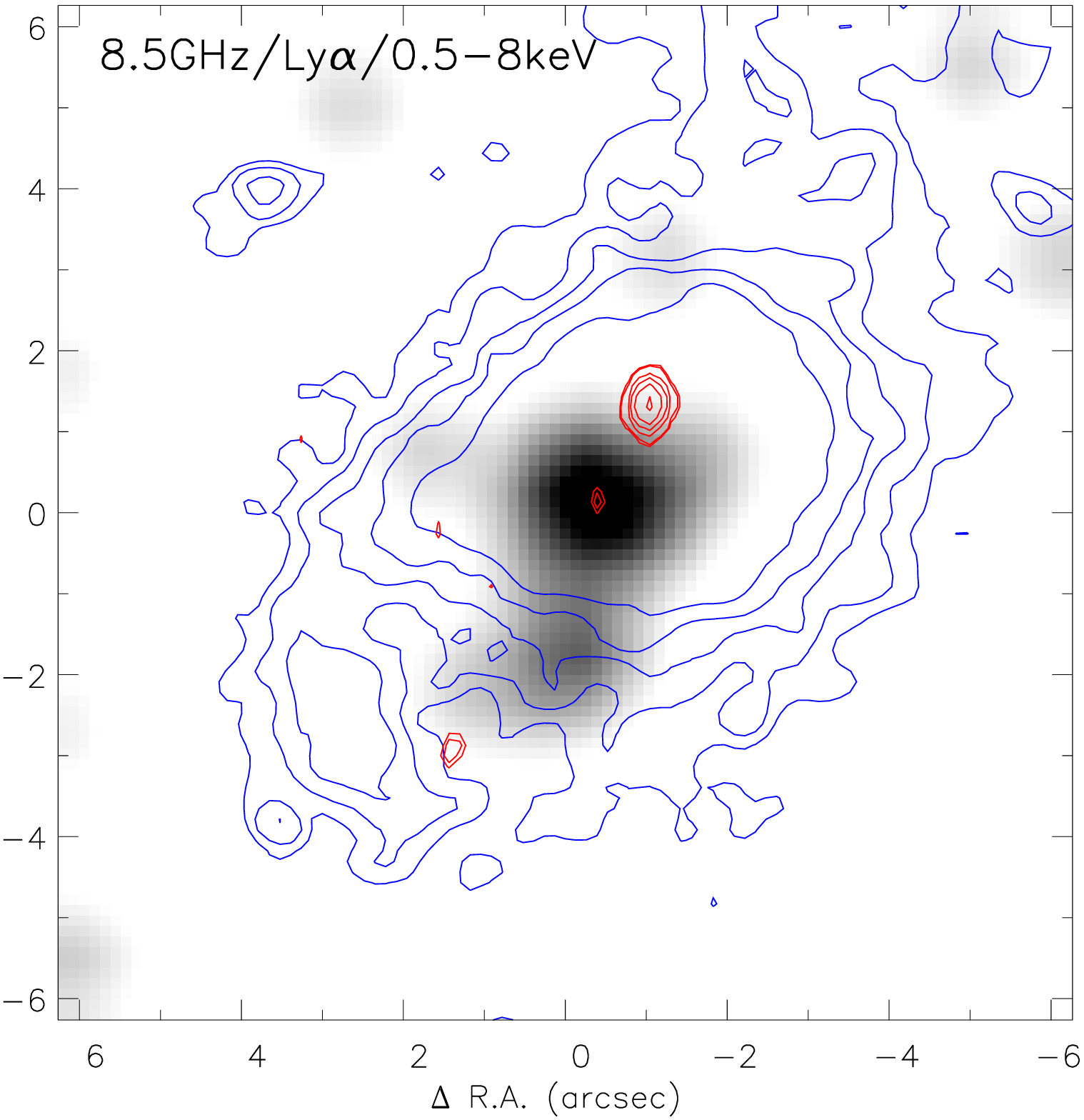,width=1.9in,angle=0}\hspace*{0.2cm}\psfig{file=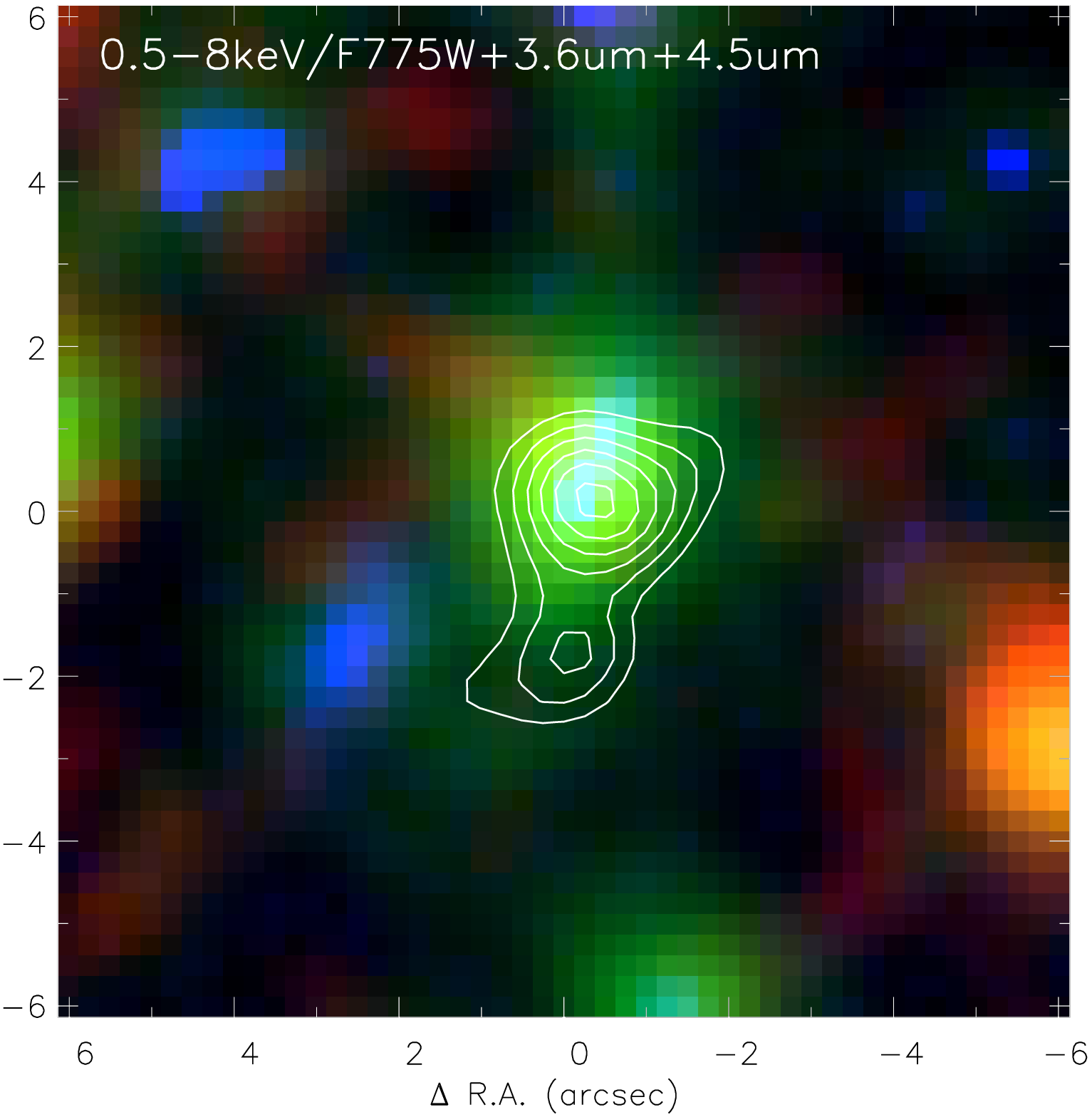,width=1.9in,angle=0}}
  \hspace*{-0.7cm}
\caption{\small Three multi-wavelength views of the region around TN\,J1338: ({\it left}) The 1.4-GHz VLA map from de Breuck et al.\ (2004) contoured over the {\it Chandra} 0.5--8\,keV X-ray observations. This shows that the X-ray emission extends southwards from the core (which is merged with the northern lobe in this relatively low-resolution radio map) to a radius of $\sim$\,20\,kpc, which is somewhat smaller than the southern radio lobe size and the X-ray emission is also mis-aligned with the lobe axis.   The lowest contour is 4-$\sigma$ (1\,$\sigma$ is 15$\mu$Jy per beam) and the increments are factors of 10$\times$, the VLA beam is 2.3$''\times $\,1.3$''$ at a PA\,=\,0$^\circ$.  The X-ray map has been smoothed with a 1$''$ FWHM Gaussian kernel to better match the radio resolution.  ({\it centre}) The Ly$\alpha$ distribution (blue contours) and 8.3-GHz VLA map (red contours) plotted on the smoothed 0.5--8\,keV X-ray observations.  We see that the Ly$\alpha$ is both significantly more extended than the X-ray emission (especially to the north, where the Ly$\alpha$ emission extends to $\gs$\,100kpc, Venemans et al.\ 2002). The Ly$\alpha$ map is from Venemans et al.\ (2002) and we truncate the highest-surface brightness regions to better show the 8.3-GHz emission from the core and radio lobes in the central regions of this compact radio source.  ({\it right}) A ``true''-colour image of the field constructed from the {\it HST} ACS F775W and {\it Spitzer} IRAC 4.5-um and 8-$\mu$m images, with the contours showing the smoothed {\it Chandra} 0.5--8-keV emission. Note the very red (8-$\mu$m peaker) source which lies $\sim $\,50-kpc in projection to the west of TN\,J1338. Similarly very red companions are seen around other high-redshift radio galaxies which exhibit luminous far-infrared emission, suggesting a possible association with the triggering of the combined AGN and starburst activity through interactions (Ivison et al.\ 2008, 2010, 2012).}
\end{figure*}

\section{Introduction}

Deep X-ray observations with the {\it Chandra X-ray Observatory} and the {\it XMM-Newton} satellite have detected spatially extended X-ray emission around more than a dozen powerful radio sources at $z\gs $\,2 (Carilli et al.\ 2002; Scharf et al.\ 2003; Fabian et al.\ 2003, 2009; Blundell et al.\ 2006; Erlund et al.\ 2006, 2008; Johnson et al.\ 2007; Smail et al.\ 2009, 2012;  Laskar et al.\ 2010; Blundell \& Fabian 2011).  This emission can extend out to $\gs $\,0.1--1\,Mpc and usually traces the morphology of the radio emitting lobes in these systems (as well as more crudely the morphology of the giant Ly$\alpha$ halos often found around these galaxies, e.g.\ Scharf et al.\ 2003).  The spatial coincidence of the X-ray emission with the radio lobes in most sources, along with its typical photon index of $\Gamma_{\rm eff}\sim$\,2 (Smail et al.\ 2012, hereafter S12), means that this emission is usually interpreted as arising from Inverse Compton (IC) scattering of CMB photons by relativistic electrons with $\gamma\sim $\,1000 in the radio lobes (e.g.\ Carilli et al.\ 2002).

At high redshift, $z\gs $\,1, the primary source of photons for the IC emission from radio galaxies is expected to be the Cosmic Microwave Background (CMB), due to its ubiquitous nature and its strongly increasing energy density at higher redshifts, $\rho_{\rm CMB} \propto (1+z)^4$ (Jones 1965; Harris \& Grindlay 1979; Nath 2010).    However, at lower redshifts, $z\ls$\,1, it has been shown that far-infrared (IR) photons from nuclear starbursts and/or hidden QSOs in radio galaxies can also provide an important contribution to their X-ray emission through IC scattering (Brunetti et al.\ 1997, 1999; Ostorero et al.\ 2010). Inverse Compton scattering of locally-produced far-infrared photons from dusty starbursts has also been suggested as a component of the X-ray emission from low-redshift ultraluminous infrared galaxies (Colbert et al.\ 1994; Moran et al.\ 1999).  Such a far-infrared-driven component of any IC X-ray emission may also occur in the more distant sources, especially because many higher-redshift radio galaxies are also luminous far-infrared  emitters, with characteristic luminosities which rapidly increase with redshift, $L_{\rm IR}\propto (1+z)^3$ (Archibald et al.\ 2001).  As Scharf et al.\ (2003) showed for 4C\,41.17, a radio galaxy with $L_{\rm IR}\sim$\,10$^{13}\,L_\odot$ at $z$\,$=$\,3.8, such extreme far-infrared luminosities mean that the far-infrared photons from the dust-obscured starburst can potentially exceed the CMB photon density on scales of $\ls $\,100\,kpc.  Thus it is possible that the luminous IC X-ray emission around compact radio sources at very high redshifts is in fact arising from a combination of two locally-produced components: $\gamma\sim $\,100 electrons in the radio lobes and far-infrared photons ($\lambda \sim $\,100\,$\mu$m) from the intense starburst occuring within these systems.  The resulting IC X-ray emission is then sufficient to ionise the gas around these galaxies, helping to form the extended Ly$\alpha$ halos which are frequently observed (e.g.\ van Breugel et al.\ 1998).  As noted by S12, this process would represent a highly effective feedback mechanism, with the AGN and starburst activity in these  galaxies combining to boost their influence on cooling gas surrounding them, and in doing so affecting evolution of the most massive galaxies seen in the local Universe.  However, thus far, much of the evidence for the role of starburst-derived far-infrared photons in the IC X-ray emission around high-redshift radio sources is circumstantial (see the discussion in S12). Hence further observations are essential to test the claimed links between far-infrared and IC emission and between IC X-ray emission and the formation of extended Ly$\alpha$ halos.

In this letter we analyse an 80-ks archival {\it Chandra} observation of TN\,J1338$-$1942, a powerful radio galaxy at $z=$\,4.11, which is a luminous far-infrared source and exhibits a striking extended Ly$\alpha$ halo.  We detect faint X-ray emission over a $\sim$\,30-kpc region which is roughly aligned with the radio lobes of this asymmetric radio source.  This is one of the highest redshift detections of X-ray emission around a radio source and the highest redshift detection for a radio galaxy (extended IC X-ray emission has been reported around radio-loud Quasars at $z=$\,4.3 and 4.7, Siemiginowska et al.\ 2003; Yuan et al.\ 2003; Cheung 2004; Cheung et al.\ 2012). As such this system provides a potential test of the competing sources of IC production photons: either the CMB or local far-infrared photons, and their influence on the gas surrounding the radio galaxy.

For our analysis we adopt a cosmology with $\Omega_{\rm m}=$\,0.27, $\Omega_\Lambda=$\,0.73 and $H_0=$\,71\,km\,s$^{-1}$\,Mpc$^{-1}$, giving an angular scale of 7.0\,kpc\,arcsec$^{-1}$ at $z=$\,4.11, a luminosity distance of 37.7\,Gpc and an age of the Universe at this epoch of 1.5\,Gyrs.

%
%
\begin{figure*}
  \centerline{\psfig{file=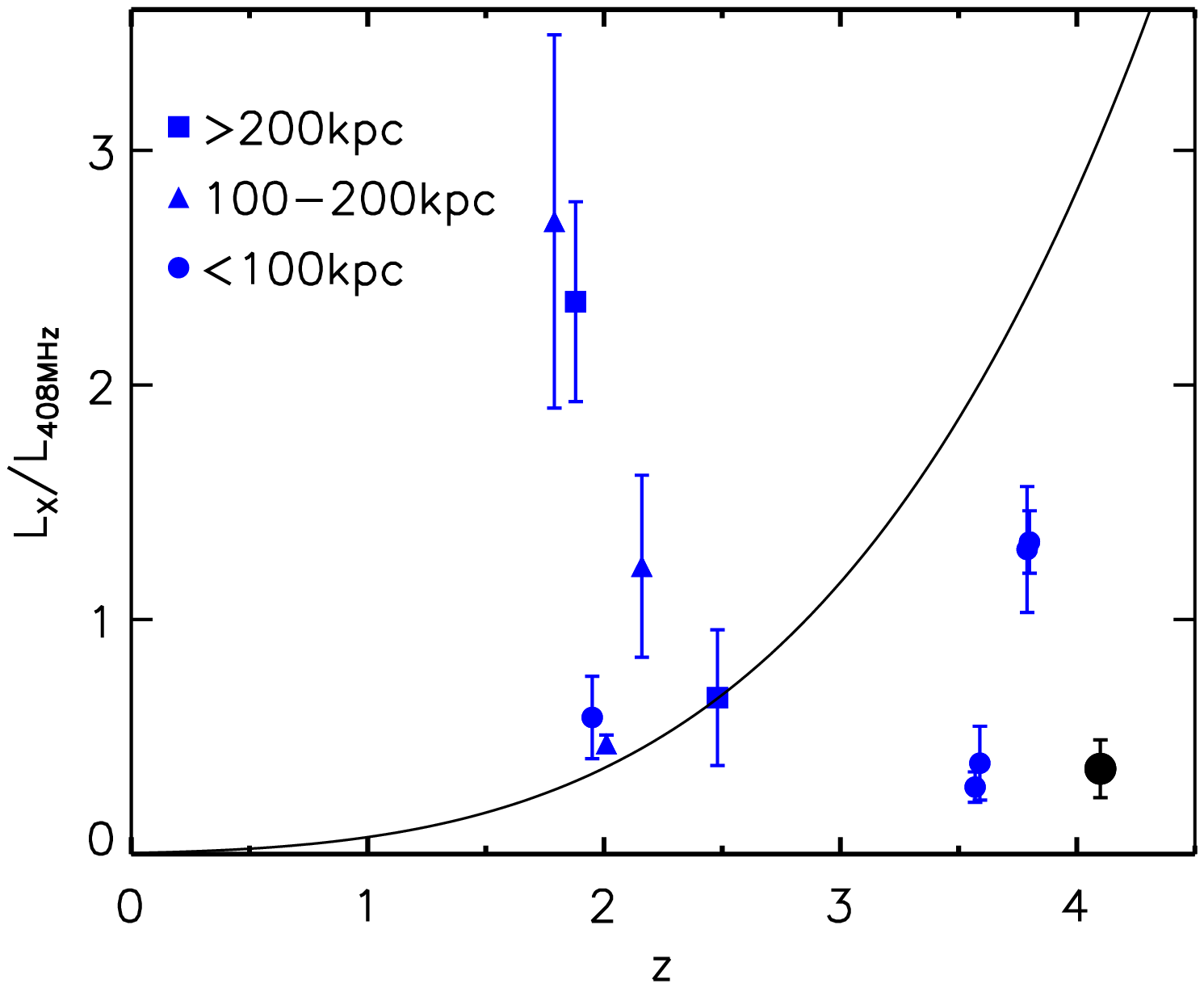,width=3.6in,angle=0}\hspace*{-1.cm}   \psfig{file=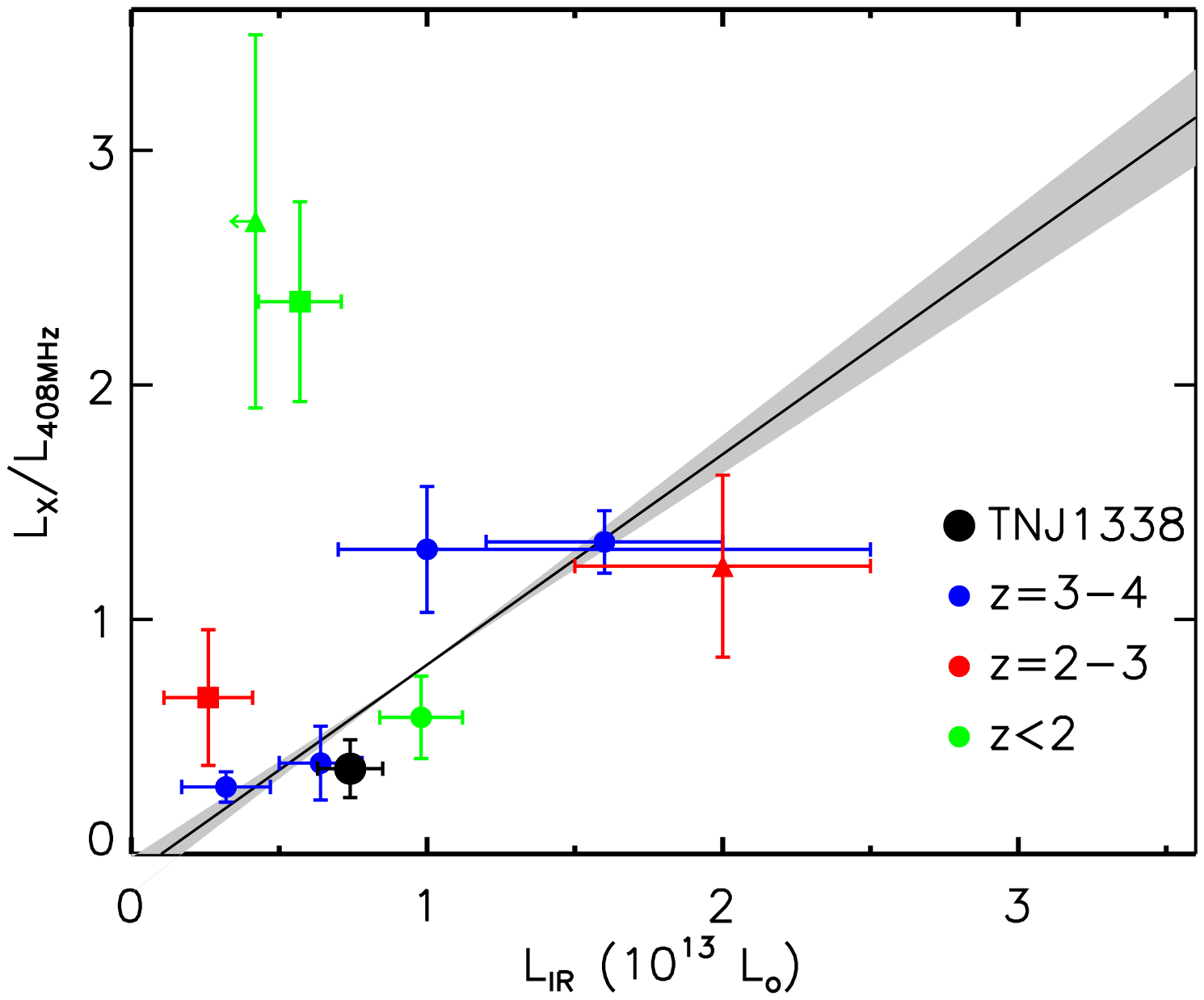,width=3.6in,angle=0}}
\caption{\small Two plots showing the variation of the X-ray to radio luminosity ratio, $L_{\rm X}$\,/\,$L_{\rm 408 MHz}$, of the IC emission around high-redshift radio galaxies as a function of either their redshifts (left-hand panel) or far-infrared luminosities (right-hand panel), using the literature compilation and data from S12.  We differentiate the radio galaxies in terms of their radio lobe lengths (see S12) and in the right-hand panel, also in terms of redshift.  In the left-hand panel we plot a toy-model track illustrating the behaviour expected for a fixed-energy bath of relativistic electrons if the IC emission is being driven by the CMB photon field, whose density rises as $(1+z)^4$. As can be seen, the data from S12 shows little evidence of such a trend and the inclusion of the new highest redshift detection for TN\,J1338, with a low $L_{\rm X}$\,/\,$L_{\rm 408 MHz}$ ratio, does not change this conclusion.  In contrast, adding TN\,J1338 to the data plotted on the right-hand panel does appear to strengthen the weak correlation which S12 claimed between the IC emission and the total infrared luminosity of the radio galaxies, especially if we remove the larger radio sources.  We plot a linear fit (1-$\sigma$ limits shown as the grey region), derived from a Monte Carlo median absolute deviation fitting routine, to the six radio sources with lobe lengths of $<$\,100\,kpc, obtaining a gradient of 0.92\,$\pm$\,0.18, consistent with unity.   These plots are adapted from S12.  }
\end{figure*}

\section{Observations and Analysis}

The radio source  TN\,J1338$-$1942 (R.A.: 13\,38\,26.11, Dec.: $-$19\,42\,32.0, J2000, hereafter TN\,J1338) was reported by de Breuck et al.\ (1999) as having redshift $z=$\,4.11.  This makes it one of the highest redshift radio galaxies known in the southern hemisphere and as a result it has been the subject of extensive multi-wavelength studies (Pentericci et al.\ 2000; Venemans et al.\ 2002; Reuland et al.\ 2004; de Breuck et al.\ 2004; Zirm et al.\ 2005; Intema et al.\ 2006; Overzier et al.\ 2008, 2009).  These studies have demonstrated that TN\,J1338 is a massive, far-infrared luminous galaxy surrounded by a Ly$\alpha$ halo with an extent of 150\,kpc\,$\times$\,40\,kpc and that it resides in the core of an over-dense structure of Ly$\alpha$ emitters and Lyman-break galaxies which have formed just $\sim$\,1.5\,Gyrs after the Big Bang.  The radio galaxy is compact, with a lobe diameter of $\sim $\,35\,kpc and a very asymmetric distribution, with the brighter northern lobe lying close to the faint core emission (Fig.~1).

Using the 850-$\mu$m flux of 6.9\,$\pm$\,1.1\,mJy from Reuland et al.\ (2004), along with their 450-$\mu$m limit, we estimate a 8--1000\,$\mu$m infrared luminosity of (7.4\,$\pm$\,1.1)\,$\times$\,10$^{12}$\,L$_\odot$ assuming a modified black body spectral energy distribution with $\beta=$\,1.5 and a characteristic dust temperature of $T_{\rm d}=$\,45\,K (see S12).  We also determine a 408-MHz luminosity of (6.5\,$\pm$\,0.4)\,$\times$\,10$^{44}$\,erg\,s$^{-1}$ from the observed 365-MHz flux of 0.72\,$\pm$\,0.14\,Jy, assuming a radio spectrum with frequency, $\nu$, which is proportional to $\nu^{\alpha}$ with $\alpha\sim -1$ (the average spectral index between 74\,MHz and 1.4\,GHz is $-0.97\pm 0.06$, Cohen et al.\ 2007; Condon et al.\ 1998).

The field of TN\,J1338 was observed for 82.4\,ks with {\it Chandra} during 2005 in {\sc faint} mode using the ACIS-S with the target placed on the nominal aim-point on the back-illuminated S3 chip.  The target was observed on three occasions for 32.4\,ks on 2005 August 29 (ObsID: 5735), 25.2\,ks on 2005 August 31 (ObsID: 6367) and a further 24.8\,ks on 2005 September 3 (ObsID: 6368).  We retrieved these data from the {\it Chandra} archive and analysed them in the same manner as Scharf et al.\ (2003) and S12. We used version 7.6.11 of the {\it Chandra} X-ray Center pipeline software for the initial data processing.  We derived  light-curves for all three observing sets, having masked all bright sources in the field, and found no significant flaring events in any of the datasets.  We then register each set of observations by running {\sc wavdetect} to generate a list of sources and matching these to the archival 4.5-$\mu$m {\it Spitzer Space Telescope} IRAC images of the field (Fig.~1) which are aligned to FK5. In this manner we confirm that the absolute astrometry of the resulting image is $\sim $\,0.4$''$. The final analysis step was then to merge the observations and construct images and exposure maps using the standard {\it ASCA} grade set ({\it ASCA} grades 0, 2, 3, 4, 6) for three bands: 0.5--8.0\,keV (full-band), 0.5--2.0\,keV (soft-band) and 2--8\,keV (hard-band).

The resulting effective on-source exposure time for the combined observations was 77.5\,ks and, after masking the emission from the core of the radio source (Fig.~1) by interpolation,  we measure a total background-corrected count of 13$^{+4}_{-3}$\,cts in the 0.5--8-keV band within a 30-kpc diameter aperture (4$''$) centered on the radio galaxy nucleus. The appropriate background was determined from randomly-placed apertures across the field.  This corresponds to a net count rate of (1.7\,$\pm$\,0.5)\,$\times 10^{-4}$\,cts\,s$^{-1}$.  Converting this count rate we determine an observed, unabsorbed 0.5--8-keV flux of (1.3\,$\pm$\,0.4)\,$\times$\,10$^{-15}$\,erg\,s$^{-1}$, assuming $\Gamma_{\rm eff}=$\,2 as expected for IC emission\footnote{The weighted mean photon index for the IC emission in a sample of 10 high-redshift radio sources derived by S12 is 1.98\,$\pm$\,0.07.} and a Galactic H{\sc i} column density  of 7.1\,$\times$\,10$^{20}$\,cm$^{-2}$. The corresponding observed-frame 0.5--8-keV luminosity is (2.2\,$\pm$\,0.7)\,$\times$\,10$^{44}$\,erg\,s$^{-1}$.  We estimate a 2--8\,keV/0.5--2\,keV flux ratio of 1.1\,$\pm$\,0.3, indicating a photon index for the emission of $\Gamma_{\rm eff}\sim 1$, with a 30\% uncertainty. Adopting this photon index to derive the observed flux would increase the estimated flux by $\sim$\,60\% and we discuss the influence of this on our results below.

\section{Results}

We show the smoothed full-band (0.5--8\,keV) X-ray map of TN\,J1338 in Fig.~1.  This figure also compares the X-ray emission to the radio and optical/mid-infrared morphology of the radio galaxy, using the 1.4-GHz radio map from de Breuck et al.\ (2004), a 8.5-GHz radio map retrieved from the  VLA archive and  archival {\it Spitzer Space Telescope} IRAC mid-infrared imaging and {\it Hubble Space Telescope} ({\it HST}) F775W ACS optical imaging (the latter has been smoothed to match the resolution of the IRAC imaging data with which it is combined). Finally, Fig.~1 also illustrates the relationship of the extended X-ray emission to the morphology of the Ly$\alpha$ halo around TN\,J1338 (Venemans et al.\ 2002).

As Fig.~1 shows, we detect X-ray emission coincident with the core of the radio galaxy, which is merged with the northern radio lobe in the relatively low-resolution VLA 1.4-GHz map (but resolved at 8.5-GHz).  In addition we see very faint extended X-ray emission around TN\,J1338 extending  out to a radius of $\sim$\,20\,kpc and aligned within $\sim$\,20$^\circ$ with the axis of the radio lobes.   The brightest X-ray emission lies to the south of the core and we estimate the likelihood that this represents a chance alignment of an unrelated X-ray source as 0.001, which along with the absence of a counterpart in the optical or mid-infrared (Fig.~1) suggests that this extended X-ray emission is associated with the radio source.  

The X-ray emission could arise from thermal bremsstrahlung emission from gas in a deep potential well (cluster) around the radio galaxy.  However, a cluster with  such a high X-ray luminosity, $\sim $\,2\,$\times$\,10$^{44}$\,erg\,s$^{-1}$ (60\% higher if we use a photon index of $\Gamma_{\rm eff}\sim 1$) similar to Coma, would be, by a large margin, the highest redshift luminous X-ray cluster: at nearly three times the redshift of the current highest-redshift, X-ray-confirmed clusters, at $z\sim $\,1.5.  We view this as unlikely and so discard this explanation, although it remains possible that this is thermal X-ray emission driven by the growth of the radio source into the clumpy intergalactic medium surrounding the galaxy. Although comparably extended X-ray emitting jets have been found around some high-redshift radio-loud quasars (e.g.\ Schwartz, et al.\ 2000; Siemiginowska et al.\ 2007), we rule out a synchrotron origin for the X-ray emission as there are no compact structures (resembling the shocks of jet knots and hotspots) in the GHz-frequency radio maps (Fig.~1).

Instead we suggest that this extended X-ray emission arises from the inverse Compton scattering of lower-energy photons by relativistic electrons associated with lobes of synchrotron plasma that have (recently or previously) been fuelled by the collimated jets from the radio galaxy central engine.   This explanation is the same as has been proposed for extended X-ray emission around similarly luminous radio galaxies at $z\sim$\,2--3.8 (see \S1), although, we note that in this example the X-ray emission is not well-aligned with the radio lobe (as in the case of e.g.\ 3C\,294, Erlund et al.\ 2008) and in addition it extends to a shorter distance than the lobe (as in the case of e.g.\ 6C\,0905, Blundell et al.\ 2006). However, this may simply reflect the challenge of detecting faint emission from relic-synchrotron plasma around this very distant source and we return to this point in the discussion.

One particularly noteworthy feature of TN\,J1338 is the small size of the radio source and its asymmetric morphology (best seen in the higher-resolution 8.5-GHz map --- Fig\,1, see also de Breuck et al.\ 1999).  TN\,J1338 is more compact than any of the other high-redshift radio galaxies which exhibit IC X-ray emission in S12, likely reflecting the youth of the radio source (Blundell \& Rawlings 1999).  High asymmetries are seen more frequently in compact, young sources (e.g.\ Saikia et al.\ 2002) and this asymmetry may have an environmental origin (McCarthy et al.\ 1991), although orientation and beaming effects are also possible. 

\section{Discussion}

We propose that the extended X-ray emission around the $z=4.11$ powerful radio galaxy TN\,J1338 arises from IC emission and we now discuss the insights this observation can provide into the photon and electron populations on which the IC process depends.

\subsection{Which seed photons are up-scattered by the relativistic electrons?}

In principle, there are two candidate seed photon fields that could be playing a role in producing  IC scattered X-ray emission:  CMB photons and far-infrared photons from the starbursts within the host galaxy.   The former would require relativistic electrons with Lorentz factors of $\gamma \sim 1000$ while the latter would require electrons with Lorentz factors of $\gamma \sim 100$ to produce KeV X-ray emission.  

To assess the prevalence of these two contributions, we first follow S12 and consider the variation in $L_{\rm X}$\,/\,$L_{\rm 408\,MHz}$ with redshift, which in a naive model of a non-evolving radio galaxy should depend only on the photon density and the magnetic field in the lobes.    If the X-ray emission were predominantly driven by an increasing CMB photon field, this ratio would increase with redshift (in the absence of systematic evolution in the magnetic fields or electron populations in radio galaxies with similar powers, sizes and ages).    In Fig.~2 we plot $L_{\rm X}$\,/\,$L_{\rm 408\,MHz}$  versus redshift and include our new observations of TN\,J1338 along with the sample of IC-detected high-redshift radio galaxies compiled by S12. Adding TN\,J1338, at the highest redshift, to the S12-compilation does not change their earlier conclusion: that there is no evidence of a correlation between $L_{\rm X}$\,/\,$L_{\rm 408\,MHz}$ and redshift (as expected from the analysis of Mocz et al.\ 2011).   This allows us to reject a very simple model where the IC emission is driven by the CMB and all the sources have comparable powers and ages. We also see that if we separate the sources on the basis of lobe size, in the expectation that the X-ray emission in the very largest systems cannot feasibly be driven by far-infrared emission from the radio galaxy (Laskar et al.\ 2010), that there is still no obvious trend with redshift. We conclude that either the CMB-driven component of the IC emission is not dominant, or if it is, that any evolving contribution to the IC scattering from CMB photons is being masked by differences in the electron populations and magnetic field strengths of the lobes in different sources (e.g.\ Mocz et al.\ 2011) at the different epochs in their life cycles at which we observe them.

In Fig.~2 we also plot $L_{\rm X}$\,/\,$L_{\rm 408MHz}$ against $L_{\rm IR}$ to discern any potential contribution from the far-infrared emission from the radio galaxy to the IC emission.  In contrast to the plot versus redshift, this shows a modest correlation, which is more obvious in the smaller, higher-redshift radio galaxies.   If we fit a linear relation to the more compact radio sources, those with $<$\,100-kpc diameter lobes, we derive a gradient of 0.92\,$\pm$\,0.18, consistent with unity.   The median absolute dispersion around the best-fit linear trend reduces by $\sim$\,35 per cent when TN\,J1338 is included in the fit. We also use a simple Monte Carlo simulation to assess the significance of the correlation, finding a $\sim 0.5$ per cent chance that the observed correlation of L$_{\rm X}$\,/\,L$_{\rm 408MHz}$ and L$_{\rm IR}$ arises by random chance. 
As noted earlier, our constraint on the photon index of the X-ray emission around TN\,J1338 is weak, if we instead adopt $\Gamma_{\rm eff}\sim 1$ then the ratio of $L_{\rm X}$\,/\,$L_{\rm 408MHz}$ will increase by $\sim 60$\%, but this has no significant effect on the correlation in Fig.~2. As suggested by S12, we therefore conclude that in the majority of more compact high-redshift radio galaxies, their luminous, dusty starbursts may be a significant source of the photons driving their IC emission.

\subsection{What is the relationship of the IC X-ray, radio and Ly$\alpha$ emission?}

A comparison of the structures of the radio, X-ray and Ly$\alpha$ emission around TN\,J1338 shows up some striking differences, especially when compared to other previously studied $z\gs 2$ radio galaxies with IC X-ray emission (e.g.\ Carilli et al.\ 2002; Scharf et al.\ 2003; Blundell et al 2006; Johnson et al.\ 2007; Erlund et al 2006, 2008; Smail et al.\ 2009, 2012) and Ly$\alpha$ halos (Pentericci et al.\ 1997; Knopp \& Chambers 1997; van Breugel et al.\ 1998). These three emission components usually have similar size-scales, but they can show morphological anti-correlations in their detailed structure.  For example the small-scale IC X-ray structure around the $z=$\,3.8 radio galaxy 4C\,60.07 appears to be anti-correlated with the structure of its Ly$\alpha$ halo (Smail et al.\ 2009) and a similar situation appears to hold for 4C\,23.56 at $z=$\,2.48 (compare Knopp \& Chambers 1997 and Blundell \& Fabian 2011).  

We also see only the broadest correlation between the X-ray and radio structures in TN\,J1338 (Fig.~1): the misalignment and smaller extent of the X-ray  compared to the GHz radio emission this indicates that, if the X-ray emission is IC-driven, then the  current (GHz-radiating) synchrotron-plasma lobes are not the source  of the electrons responsible for the IC scattering.   This then requires that  {\it relic} synchrotron plasma, no longer radiant at GHz-frequencies, provides the relevant reservoir of relativistic electrons for the IC scattering process, such as predicted by episodic models of jet ejection in quasars and radio galaxies (Nipoti et al.\ 2005; Blundell \& Fabian 2011).  If this speculation is correct then it suggests in turn that lower-energy ($\gamma \sim $\,100) rather than higher-energy ($\gamma \sim $\,1000) electrons are plentiful, which is in accordance with the picture that far-infrared photons dominate the upscattering process (rather than CMB photons which must scatter from the rarer, higher-energy  electrons with $\gamma \sim $\,1000).

There are also mismatches  between the compact scale of the radio and X-ray emission and the much more extended northern Ly$\alpha$ ``lobe'' which spans over 15$''$ (100\,kpc) from the radio core (Fig.~1; Venemans et al.\ 2002).  We have searched for extended X-ray emission associated with the Ly$\alpha$ lobe in the hard and soft band {\it Chandra} images and find no emission above the background in this region. Given the short cooling timescale expected for the Ly$\alpha$ halo (Geach et al.\ 2009), it appears unfeasible that it represents cooling associated with the current radio lobe or its direct IC X-ray emission, as proposed for other high-redshift radio sources (Scharf et al.\ 2003; S12).     Very speculatively, we suggest that the Ly$\alpha$ halo could represent cooling associated with heating by IC scattering from an even older synchrotron plasma produced by an earlier phase of activity, whose Lorentz factors of $\ls 100$ could upscatter CMB photons into UV photons.  Long-baseline ($\gs$\,1000-km) low-frequency radio observations, similar to those made possible by the most extended configurations of LOFAR, including LOFAR-UK, would be needed to search for this electron population and compare its  distribution to that of the Ly$\alpha$  emission to test this  suggestion.   Until such observations are available, the lack of correspondence between  the Ly$\alpha$ halo and the radio and X-ray emission in this system remains a puzzle.

\section{Conclusions}

We analyse a deep archival {\it Chandra} X-ray observation of the powerful radio galaxy TN\,J1338$-$1942 at $z=$\,4.11 and detect faint X-ray emission extending over a $\sim$\,30-kpc region around the radio core.   We propose that the X-ray emission most likely arises from IC scattering, by relativistic electrons in the synchrotron plasma lobes, of sub-millimetre photons from the CMB or far-infrared photons from the dusty, starburst in this galaxy.  This is one of highest redshift detections of IC X-ray emission around a radio source and the highest known for a radio galaxy.

We compare the relative strength of the X-ray and radio emission in this system to other high-redshift, IC-emitting radio sources from the literature. On the basis of this and other evidence we find support for the claim that the IC emission in the majority of these systems (those with radio lobes with sizes less than $\sim$\,100\,kpc) may be enhanced by the far-infrared photons from the dusty starbursts occuring in these active high-redshift galaxies.

It has been proposed that the luminous IC X-ray emission around compact, high-redshift far-infrared-luminous radio galaxies represents a significant feedback process (Scharf et al.\ 2003; S12).  The particle and photon emission from the lobes and starburst in these massive, radio-loud composite AGN and starbursts, would combine to produce intense X-ray emission distributed across 10--100\,kpc scales, encompassing the densest regions of the gas reservoir surrounding these galaxies.  The heating from these X-rays can help create an ionised halo of gas and as a result reduces the amount of material available to cool onto the galaxy and subsequently form stars.  However, we stress that in TN\,J1338 we see little correspondence between the very extended Ly$\alpha$ cooling halo around the galaxy and its more compact KeV X-ray and GHz radio emission, which argues for a more complex origin for this feature.  

\section*{Acknowledgements}
We thank the anonymous referee for their constructive comments on this paper and we also thank Bret Lehmer for help and Carlos de Breuck for generously sharing his 1.4-GHz VLA map. IRS acknowledges support from STFC, a Leverhulme Fellowship, the ERC Advanced Programme {\sc DustyGal} and a Royal Society/Wolfson Merit Award.  This work has used data from the NASA Extragalactic Database (NED), and from the NRAO VLA, {\it HST} and {\it Spitzer} archives.

\end{document}